\documentclass[final]{svjour2}
\usepackage{graphicx}
\usepackage{rotating}
\usepackage{amssymb}
\usepackage{mathptmx}
\usepackage{bm}
\usepackage[numbers]{natbib}
\makeatletter
\journalname{Journal of Low Temperature Physics}

\bibpunct{}{}{,}{s}{}{,}

\begin{document}
\bibliographystyle{apsrev4-1}
\newcommand{\hdblarrow}{H\makebox[0.9ex][l]{$\downdownarrows$}-}
\title{Spin superfluidity, coherent spin precession, and magnon BEC}

\author{E.B. Sonin$^1$}

\institute{1:Racah Institute of Physics, Hebrew University of Jerusalem, Givat Ram, Jerusalem 91904, Israel\\
\email{sonin@cc.huji.ac.il}
}

\date{\today}

\maketitle

\keywords{Spin superfluidity, coherent spin precession, and magnon BEC}

\begin{abstract}

Spin superfluidity, coherent spin precession, and magnon BEC are intensively  investigated theoretically and experimentally nowadays. Meanwhile, clear definition and differentiation between these  related phenomena is needed. It is argued that spin stiffness, which leads to existence of coherent spin precession and dissipationless spin supercurrents, is a necessary but not sufficient condition for spin superfluidity. The latter is defined as  a possibility of spin transport on macroscopical distances with sufficiently large spin supercurrents. This possibility is realized at special topology of the magnetic-order-parameter space, such as, e.g., that in easy-plane antiferromagnets. It is argued that an arbitrarily chosen formal criterion for the existence of magnon BEC has no connection with conditions for observation of  macroscopic dissipationless spin transport.

PACS numbers: 75.76.+j,47.37.+q,67.30.hj
\end{abstract}

\section{Introduction}
\label{intro}

The problem of dissipationless spin transport  also called spin superfluidity has occupied minds of condensed matter physicists for decades. A similar phenomenon of  superfluidity of electron-hole pairs was discussed from 60s\cite{KozMax}. The concept of spin superfluidity was exploited in the attempt to explain experiments demonstrating unusually fast spin relaxation in $^3$He-A \cite{Vuo,CorOsh}. 
These ideas were confronted by the argument that the absence of the strict conservation laws for electron-hole pairs and spin rules out any analogy with mass superfluidity. Nevertheless, it was demonstrated, first for  electron-hole pairs\cite{LY,Sh,ES-77}  and then for spin\cite{ES-78a,ES-78b,ES-82}, that for weak violation of the conservation law analogy with mass superfluidity is still possible. 
Later on  intensive theoretical and experimental work on spin superfluidity in superfluid $^3$He-B was done. \cite{HeBex,Bun} Nowadays  we observe a growing interest to superfluid  spin transport in connection with work on spintronics, where transport of spin with minimal losses is crucial.  

 It seems useful to have a glance on the current status of the field. The intention is to discuss mostly concepts without entering into details, which a reader can find in recent reviews.\cite{Adv,BunV} Since from the very beginning of the  theory of superfluidity the relation between superfluidity and Bose--Einstein Condensate (BEC) was permanently in the focus of attention, discussing spin superfluidity one cannot avoid to consider its relation to the concepts of coherent states (coherent spin precession, in particular) and magnon BEC. 
 
\section{What is  superfluidity and  superfluid transport?}

Sometimes the term ``superfluidity'' is used in the literature to cover a broad range of phenomena, which have been observed in superfluid $^4$He and $^3$He, Bose-Einstein condensates of cold atoms, and, in the broader sense of this term, in superconductors. We prefer to define superfluidity  only as a possibility to transport a physical quantity (mass, charge, spin, ...) without dissipation (or, in more accurate terms, with suppressed dissipation). Exactly this phenomenon  gave a rise to the terms ``superconductivity'' and ``superfluidity'', discovered  nearly 100 years  and 70 years ago respectively. 

The essence of the transition to the superfluid or superconducting state is that below the critical temperature the complex order parameter $\psi =|\psi| e^{i\varphi}$, which has a meaning of the wave function of the bosons or the fermion Cooper pairs, emerges as an additional macroscopical variable of the liquid.   For the sake of simplicity, we restrict ourselves to the case of a neutral superfluid at zero temperature. The theory of superfluidity tells that the order parameter $\psi$ determines the particle density $n=|\psi|^2$   and the superfluid velocity of the liquid is given by the standard quantum-mechanical expression
\begin{equation}
\bm v_s = -i {\hbar \over 2 m |\psi|^2}(\psi^* \bm \nabla \psi-\psi \bm \nabla \psi^*)= {\hbar \over m} \bm \nabla \varphi.
         \end{equation}
Thus the velocity is a gradient of a scalar, and any flow is potential.

 \begin{figure}
\begin{center}
\includegraphics[%
  width=0.9\linewidth,
  keepaspectratio]{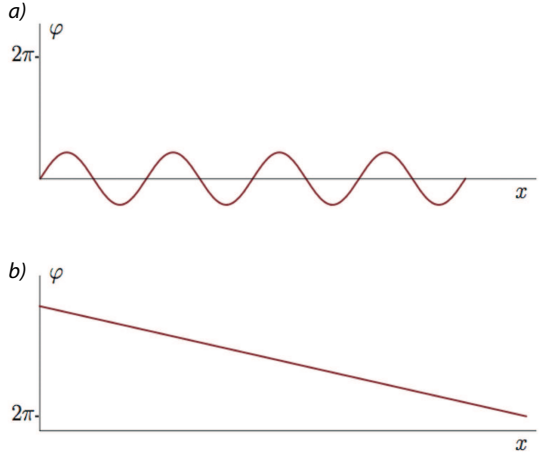}
\end{center}
\caption{(Color online) Phase (inplane rotation angle)  variation at the presence of mass (spin) supercurrents. a) Oscillations in a sound (spin) wave). b) Stationary mass (spin) supercurrent.}
\label{fig1}
\end{figure}

An elementary collective mode of an ideal liquid is a sound wave. In a sound wave the phase varies in space, i.e., the wave is accompanied by mass supercurrents (Fig.~\ref{fig1}a). An amplitude of the time and space dependent phase variation is small, and currents transport mass on distances of the order of the wavelength, which is not a macroscopic transport yet. A really superfluid transport on macroscopic distances is related with stationary solutions of the hydrodynamic equations corresponding to finite constant currents  (current states). In the current state the phase performs a large number of full 2$\pi$-rotations along streamlines of the current (Fig.~\ref{fig1}b). 

According to the Landau criterion, the current state is stable as far as \emph{any elementary excitation} of the Bose-liquid in the laboratory frame has a positive energy and  its creation requires an energy input. This yields the Landau critical velocity:
$v_L =\mbox{min}\{\varepsilon(\bm p)/p\}$, where $\varepsilon(\bm p)$ is an energy of an excitation with momentum $\bm p$.
In superfluid $^4$He elementary excitations are phonons and rotons, and the Landau critical velocity $v_L$ is determined by the roton part of the spectrum. Anyway, the supercurrent cannot be stable if the velocity  exceeds the sound velocity $u_s$. 

The Landau velocity determines stability with respect to weak perturbations (single-particle excitations). Meanwhile, a real process of supercurrent decay is realized at high velocities via motion of vortices across current streamlines (phase slips). This motion is impeded by energetic barriers which disappear when the superfluid velocity becomes of the order of $\hbar / mr_c$ where $r_c$ is a core radius. Since in the Bose liquid $r_c \sim \hbar /mu_s$ the stability with respect to phase slips yields approximately the same criterion as the Landau criterion: $v_s<u_s$. Note, however, that this is only an upper bound for the critical velocity, since the energetic barriers impeding vortex motion can be overcome by thermal activation or quantum tunneling.

\begin{figure}
\begin{center}
\includegraphics[%
  width=0.7\linewidth,
  keepaspectratio]{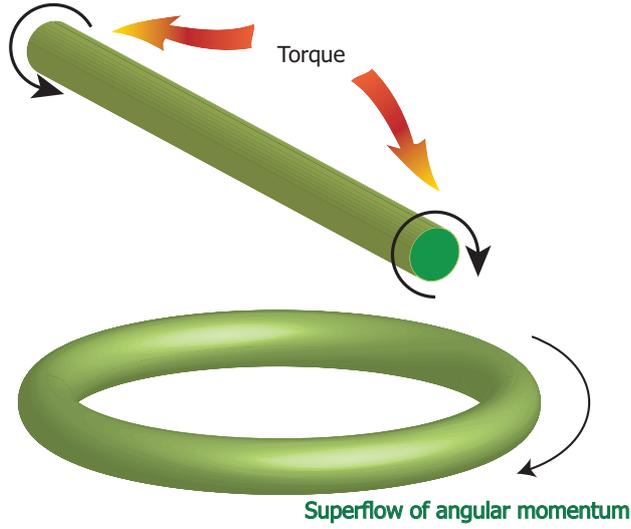}
\end{center}
\caption{(Color online) Mechanical analogue of a persistent current: A twisted elastic rod bent into a closed ring. There is a persistent angular-momentum flux around the ring.}
\label{fig3a}
\end{figure}

Before starting discussion of spin superfluid transport  it is useful to  consider a mechanical analogue of superfluid mass or spin supercurrent \cite{ES-82,Adv}. Let  us twist a long elastic rod so  that  a twisting angle at one end of the rod with respect to an opposite end  reaches values many times $2\pi$. Bending the rod into a ring and connecting the ends rigidly, one obtains a ring with a circulating persistent angular-momentum flux (Fig.~\ref{fig3a}). The intensity of the flux is proportional to the gradient of twisting angle, which plays the role of the phase gradient in the mass supercurrent or the spin-rotation-angle gradient in the spin supercurrent. The analogy with spin current is especially close because spin is also a part of the angular momentum.  The deformed state of the ring is not the ground state of the ring, but it cannot relax to the ground state via any elastic process, because it is topologically stable.  The only way to relieve the strain inside the rod is {\em plastic displacements}. This means that dislocations must move across rod cross-sections. The role of dislocations in the twisted rod is the same as the role of vortices in the mass or spin current states: In both of the cases some critical deformation (gradient) is required to switch the process  on.

\section{Spin supercurrents and spin conservation law} \label{AF}

For the sake of simplicity we have in mind a two-sublattice antiferromagnet with sublattice magnetizations $\bm M_1=\bm M_0$ and $\bm M_2=-\bm M_0$. In the equilibrium ``easy plane'' anisotropy keeps the magnetizations in the easy plane $xy$.    There is also  $n$-fold  anisotropy inside the easy plane, and the free energy  can be written as
\begin{eqnarray}
{\cal F}=\int d^3\bm R\left\{{ m_z^2\over 2\chi} + {A(\bm \nabla \varphi)^2\over 2}+K[1-\cos (n\varphi)] \right\}.
 \label{an}  \end{eqnarray}
Here $m_z$ is a small $z$ magnetization in a non-equilibrium state when the sublattice magnetizations slightly go out of the easy plane, and  the angle $\varphi$ determines orientation of the antiferromagnetic vector (staggered magnetization) $\bm L= \bm M_1-\bm M_2$ in the easy plane.  
The constant  $A$ is stiffness of the spin system determined by exchange interaction,  and
the magnetic susceptibility $\chi= M_0^2/E_A$ along the $z$ axis is determined by the uniaxial anisotropy energy $E_A$ keeping the magnetization in the plane. 
The Landau-Lifshitz equation reduces to the Hamilton equations for a pair of canonically conjugate continuous variables ``angle--angular momentum'' (analogous to the canonically conjugate pair ``coordinate--momentum''):
         \begin{eqnarray}
{d\varphi \over dt}=-\gamma { m_z\over \chi},
     \label{Ep} \end{eqnarray}
   \begin{eqnarray}
{1\over \gamma}{dm_z \over dt}=\bm \nabla \cdot \bm J^z+nK \sin(n\varphi)=- A\left[\nabla^2 \varphi -{\sin(n\varphi)\over l^2}\right] ,
 \label{EmK}      \end{eqnarray}
where $\gamma$ is the gyromagnetic ratio,  
\begin{eqnarray}
\bm J^z=-{\partial F \over \partial \bm \nabla \varphi} =-A   \bm \nabla \varphi
   \label{cur}    \end{eqnarray}
is the spin current, and the scale
 \begin{eqnarray}
l=\sqrt{ A\over nK}.
             \end{eqnarray}
determines the  thickness of domain wall separating possible domains with various $n$ directions of sublattice magnetizations.  

If the inplane anisotropy is absent ($K=0$)  the $z$ component of spin is conserved, and there is an evident analogy of Eqs.~(\ref{Ep}) and (\ref{EmK}) with the hydrodynamic equations, Eq.~(\ref{EmK})  being the continuity equation for spin. This analogy was exploited by Halperin and Hohenberg \cite{HH} in their hydrodynamic theory of spin waves. In easy-plane ferromagnets spin waves have a sound-like spectrum as in a superfluid: $\omega=c_s k$, where the spin-wave velocity is $c_s=\gamma \sqrt{A/\chi}$. Halperin and Hohenberg introduced the concept of 
 spin current, which appears in a propagating spin wave like a mass supercurrent appears in a sound wave (Fig.~\ref{fig1}a). This current transports the $z$ component of spin on distances of the order of the wavelength.
 But as well as the mass supercurrent in a sound wave, this small oscillating spin supercurrent does not lead to superfluid spin transport on macroscopical scales. Spin superfluid transport on long distances is realized in current states with magnetization rotating monotonously in the plane as shown in Fig.~\ref{fig1}b. 

First discussions of spin superfluidity\cite{Vuo,CorOsh} ignored processes violating the conservation law for the total spin. Though these processes are relativistically weak, their  effect is of principal importance and in no case can be ignored. The attention to superfluid transport   in the absence of conservation law was attracted first in connection with discussions of superfluidity of electron-hole pairs. The number of electron-hole pairs can vary due to interband transitions. As was shown by Guseinov and Keldysh \cite{GK}, interband transitions lift the degeneracy with respect to the phase of the  ``pair Bose-condensate''   and make the existence of spatially \emph{homogeneous} stationary current states impossible. This phenomenon was called ``fixation of phase''.  On the basis of it Guseinov and Keldysh concluded that no analogy with superfluidity is possible without conservation law.  At that period this stance became a common wisdom, which ruled out also spin superfluidity. Meanwhile it was shown\cite{LY,Sh,ES-77}  that although ideally uniform current states are impossible without conservation law indeed, still there are possible slightly non-uniform  electron-hole-pair-current states, which can mimic states with stationary mass supercurrents. 
This analysis was extended on spin currents \cite{ES-78a,ES-78b,ES-82}.

\begin{figure}
\begin{center}
\includegraphics[%
  width=1.00\linewidth,
  keepaspectratio]{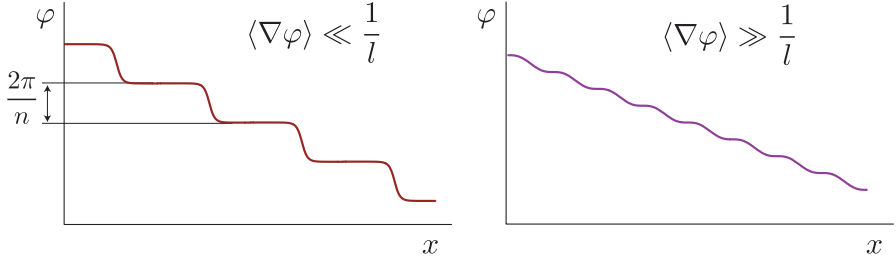}
\end{center}
\caption{(Color online) The nonuniform spin-current states with $\langle \nabla \varphi \rangle \ll 1/l$ and $\langle \nabla \varphi \rangle \gg 1/l$.}
\label{fig2}
\end{figure}

In the spin system the role of the phase is played by the angle $\varphi$ in the easy plane, and the degeneracy with respect to the angle is lifted by inplane anisotropy  $K$. Excluding $m_z$ from Eqs.~ (\ref{Ep}) and (\ref{EmK})  one obtains the sine Gordon equation for the angle $\varphi$:
  \begin{eqnarray}
{\partial ^2 \varphi \over \partial t^2}- c_s^2\left[\nabla^2 \varphi -{\sin(n\varphi)\over l^2}\right]=0 .
          \end{eqnarray}
Stationary solutions of this equation are shown in Fig.~\ref{fig2}. At small average gradients $\langle \nabla \varphi \rangle \ll 1/l$ the spin-current state is a chain of well separated domain walls of width $l$ and have no similarity with mass supercurrent states. On the other hand, at large average gradients $\langle \nabla \varphi \rangle \gg 1/l$ the spin-current state is nearly uniform mimicking the mass-supercurrent state (Fig.~\ref{fig1}).    

\section{Stability of spin-current states: Landau criterion} \label{stab}

Like in the case of mass supercurrents, the spin-current state is metastable and corresponds to a local minimum of the free energy, i.e., any transition to nearby states would require an increase of energy. This condition leads to the Landau criterion.   
In order to check current metastability, one should estimate the energy of possible small static fluctuations around the stationary current state. For this estimation, one should take into account that the stiffness constant $A$ is proportional to the squared inplane  component of the sublattice magnetization  $ M_{\perp}^2 = M_0^2-m_z^2/4$, and in the presence of large angle gradients $A$ must be replaced with $A(1-m_z^2/4M_0^2)$. So the free energy is 
\begin{eqnarray}
{\cal F}=\int d^3\bm R\left[{ m_z^2\over 2\chi} + {A(1-m_z^2/4M_0^2)(\bm \nabla \varphi)^2\over 2} \right]
\nonumber \\
=\int d^3\bm R\left[{ m_z^2\over 2}\frac{E_A-A(\bm \nabla \varphi)^2/4}{M_0^2} + {A(\bm \nabla \varphi)^2\over 2} \right].
   \end{eqnarray}
 One can see that if $\nabla \varphi$ exceeds $2\sqrt{E_A/ A}=2c_s M_0/\gamma A$ the current state is unstable with respect to the exit of $\bm M_0$ from the easy plane. This is the Landau criterion for the stability of the spin current.  

In  conventional mass superfluidity the supercurrent  is restricted only from above, by the Landau critical velocity. In contrast, in spin superfluidity (as in any other superfluidity of nonconserved quantities) more or less uniform supercurrents are also restricted from below, by supercurrents of the order of those, which exist in domain walls. Superfluidity is observable only if  the Landau critical supercurrent essentially exceeds supercurrents in domain walls.\footnote{ \citet{McD} came independently to a  similar conclusion concerning dissipationless spin transport in thin film ferromagnets.}  Like in superfluids, stability of current states is connected with topology of the order parameter space.  For isotropic antiferromagnets the space of degenerated equilibrium states is a sphere $|\bm L|=const$, whereas for an easy-plane antiferromagnet strong uniaxial anisotropy keeps the sublattice magnetizations in the easy plane reducing the order parameter space to  the equatorial circumference similar to the order parameter space in usual superfluids.

 As well as in the theory of mass superfluidity, after reaching the Landau critical gradient the current state becomes unstable with respect to large perturbations, which are \emph{magnetic vortices}.  
The    magnetic vortex energy is  determined by the expression similar to that  for a usual superfluid vortex: 
\begin{eqnarray}
\epsilon =\int d^2\bm r {A(\bm \nabla \varphi)^2\over 2} =\pi A \ln{r_m\over r_c},
    \label{vorEn}
          \end{eqnarray}
where the upper cut-off $r_m$ depends on geometry. However, the radius $r_c$ and the structure of the magnetic vortex core are determined differently  from the mass vortex.\cite{ES-78a,ES-78b} In a magnetic system the order parameter must not vanish at the vortex axis since
there is a more effective way to eliminate the singularity in the gradient energy:  an excursion of the spontaneous magnetization out of the easy plane $xy$. This would require an increase of the uniaxial anisotropy energy, which keeps $\bm M_0$ in the plane, but normally this energy is much less than the exchange energy, which keeps the order-parameter amplitude $M$ constant. Finally the core size $r_c$ is determined as a distance at which the uniaxial anisotropy energy density $E_A$ is in balance with the gradient energy  $A(\nabla \varphi)^2 \sim A/r_c^2$.  This yields $r_c \sim \sqrt{A/E_A}$.  In contrast to superfluid vortices mapping onto a plane circle, the spin vortex state can map onto one of two halves of the  sphere $|\bm L|=const$. Thus a magnetic (spin) vortex has an additional topological charge having two values $\pm 1$ \cite{Nik,Adv}.

The energy of the spin-current state with a vortex and the energy of the barrier, which blocks the phase slip,  i.e., the decay of the current, are determined similarly to the case of mass superfluidity. The barrier disappears at gradients $\nabla \varphi_0 \sim 1/r_c$, which are of the same order as the critical gradient determined from the  Landau criterion. This is a typical situation in the superfluidity theory. But sometimes the situation is more complicated (see Sec.~\ref{HeB}). 

\section{Observation of  superfluid spin transport }

\begin{figure}
\begin{center}
\includegraphics[%
  width=0.98\linewidth,
  keepaspectratio]{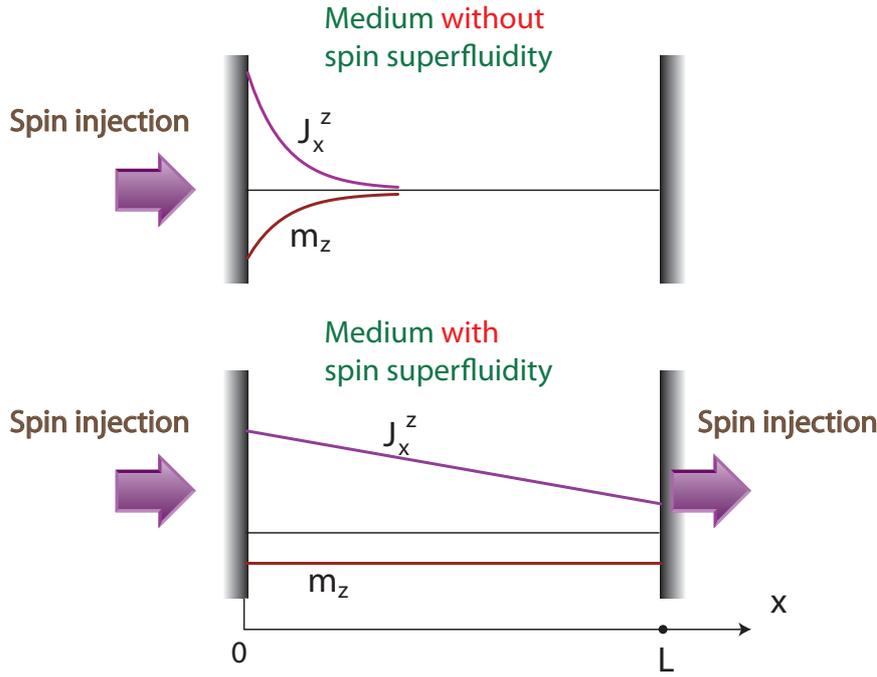}
\end{center}
\caption{(Color online) Spin injection to a spin-nonsuperfluid and a spin-superfluid medium.}
\label{fig4}
\end{figure}

Let us discuss possible demonstration of superfluid spin transport.\cite{ES-78b,ES-82} Suppose that spin is injected into a sample  at the sample boundary $x=0$ (Fig.~\ref{fig4}). The injection can be realized  either with an injection of  a spin-polarized current, or with pumping the spin with a circularly polarized microwave irradiation.  If the medium at $x>0$ cannot support superfluid spin transport,  the only way of spin propagation is spin diffusion, and both the spin current and the nonequilibrium magnetization $m_z$ exponentially decay inside the sample: $J_d^z \propto m_z \propto e^{- x/L_s}$,  where $L_s=\sqrt{D_sT_1}$ is the spin-diffusion length,  $D_s$ is the spin-diffusion coefficient,  and  $T_1$ is the time characterizing the Bloch longitudinal  relaxation, which violates the spin-conservation law.  So no spin can reach the other boundary $x=L$ of the sample provided $L \gg L_s$. 

Now let us suppose that the medium at $0<x<L$ is magnetically ordered and can support superfluid spin transport.
Neglecting  inplane anisotropy, which is justified at strong injection ($\nabla \varphi \gg 1/l$),  spin transport is described by the equations 
        \begin{eqnarray}
{d\varphi \over dt}=-\gamma { m_z\over \chi},
     \label{EpB} \end{eqnarray}
     \begin{eqnarray}
{dm_z \over dt}-\gamma\bm \nabla \cdot \bm J^z+ {m_z \over T_1}=0,
 \label{EmB}      \end{eqnarray}
with the boundary conditions for the supercurrent  $\bm J^z(0) = \bm J^z_0$ at $x=0$ and $\bm J^z(L) =- f m_z (L) $ at $x=L$. The current $J^z_0$ in the first condition is the spin-injection current, while the second boundary condition takes into account that the medium at $x>L$ is not spin-superfluid and spin injection there is possible only if some non-equilibrium magnetization $m_z(L)$ is present. The coefficient $f$ can be found by solving   the spin-diffusion equations in the medium at  $x>L$.  While  the inplane anisotropy violating the spin conservation (phase fixation) was neglected, one cannot neglect irreversible dissipative processes, which also violate the spin-conservation law. The simplest example of such a process is  
the  longitudinal spin relaxation characterized by time $T_1$. 

The stationary solution of Eqs.~(\ref{EpB}) and (\ref{EmB}) is\cite{ES-78b} 
    \begin{eqnarray}
m_z= -{\gamma  T_1 \over L+f \gamma T_1}J^z_0\approx- {\gamma  T_1 \over L}J^z_0,~~J^z(x) = J^z_0 \left(1-{x\over L+f \gamma T_1} \right) \approx J^z_0{L-x\over L}.
     \end{eqnarray}
The solution is stationary in the sense that  $\partial m_z /\partial t=0$, but slow stationary precession takes place:  $\partial \varphi /\partial t \neq 0$. We consider a non-equilibrium process (otherwise spin accumulation is impossible), which is accompanied by the precession of $\bm M_0$ in the easy plane. But the process is stationary  only if the precession angular velocity is constant in space. The condition $m_z=$const, which results from it, is similar to the condition of constant chemical potential in superfluids or electrochemical potential in superconductors in stationary processes. If this condition were not satisfied, there would be steady growth of the angle twist as is evident from Eq.~(\ref{EpB}). 

\section{Spin superfluidity in $^3$He-B} \label{HeB}

The general concept of spin superfluidity presented here is relevant  to spin superfluidity in  $^3$He-B,\cite{HeBex,Bun} but the latter  has some  features, which distinguish it from the model of spin superfluidity discussed in the previous sections. First, in contrast to what was considered earlier, observed spin-current states in the $B$ phase are dynamical nonlinear states very far from the equilibrium, which require for their support permanent pumping of energy.  Second, while the previous discussion dealt with the degree of freedom connected with the longitudinal magnetic resonance,  in the $B$ phase spin vector performs a more complicated 3D rotation, but still well described by one degree of freedom connected with the transverse magnetic resonance (nuclear magnetic resonance in the case of $^3$He).  

In the past the group, which studied spin superfluidity in  $^3$He-B, objected to some principles  of the spin-superfluidity theory, which was presented above.  First, they subscribed to common wisdom of that time that spin superfluidity is impossible without strict conservation law. Therefore, time and again they wrote in their papers that spin  superfluidity was possible only  in  $^3$He-B because the Hamiltonian describing the spin precession  in  $^3$He-B does not contain any term violating the conservation law and fixing the phase of precession.\cite{AIP,Fom91,Bun} Second,  they stressed that superfluid spin transport  in superfluid $^3$He was related with a counterflow of particles with opposite spins and ruled out spin superfluidity in solids with magnetic order   resulting from exchange interaction between localized spins (see, e.g.,  p. 92 in the review by \citet{Bun}).  In contrast, in our theory of superfluid spin transport  it does not matter whether  magnetism is connected with itinerant or localized spins.\cite{ES-78b}.   In the latest paper \citet{BunLvo} addressed  spin superfluidity in solid antiferromagnetic insulators, where there is no conservation law for spin and spin carriers are localized. One may interpret this as that Bunkov retracted his former criticism. 

Application of the Landau criterion\cite{ES-87} for spin superfluidity in  $^3$He-B was also disputed. 
Fomin \cite{Fom-88} stated that the Landau criterion is not necessary for the superfluid spin transport since emission of spin waves, which comes into play after exceeding the Landau critical gradient, is weak in the experimental conditions (see also the similar conclusion after Eq.~ (2.39) in the  review by  Bunkov \cite{Bun}). This stance confuses superfluid and  ballistic transport. If  observed  spin transport were ``dissipationless'' simply  because dissipation was weak,  it would be ballistic  rather than superfluid transport. The essence of the phenomenon of superfluidity is not the absence of sources of dissipation, but ineffectiveness of these sources due to energetic and topological reasons. The Landau criterion is an absolutely necessary condition for superfluidity. Fortunately for the superfluidity scenario in the $^3$He-B, Fomin's estimation of the role of dissipation by spin-wave emission triggered by violation of the Landau criterion was not conclusive.\cite{Adv}  The misconception concerning the role of the Landau criterion for spin superfluidity in  $^3$He-B existed up to recent days, when finally \citet{BunV}  (see their Sec.~V.H) accepted applicability of the Landau criterion for spin superfluidity. 

But another misconception concerning stability of supercurrents in $^3$He-B still remains unsettled.
As mentioned above the spin current at which stability with respect to vortex nucleation and growth is lost (i.e., the barrier for the growth disappears) is the same as that obtained from the Landau criterion. The barrier  for vortex growth in the phase-slip process vanishes at phase gradients of the order of the inverse core radius. In the first paper on the spin vortex in  $^3$He-B  the vortex core radius was estimated to be on the order of the dipole length  \cite{ES-87}, which agrees with the critical gradient from the Landau criterion.  Later Fomin \cite{Fom-88} showed that close the critical angle $104^\circ$ of precession the vortex core is determined by another much longer scale.   Since no barrier impedes vortex expansion across a channel if the gradient is on the order of $1/r_c$, the large core $r_c$   leads to the strange (from the point of view of the conventional superfluidity theory) conclusion: The instability with respect to vortex creation occurs at the phase gradients  essentially less than the Landau critical gradient.  Exactly this was stated by \citet{BunV} in  their Sec.~V.H, even though this would mean again that the Landau criterion is irrelevant.  But if in the past the Landau criterion was rejected  because it predicted a too low critical gradient now it is rejected as predicting a too high critical gradient. In reality, there is no disagreement between the critical gradient for vortex creation and that determined from the Landau criterion:
Recently it was shown \cite{ES-08,Adv} that at precession angles close to 104$^\circ$, when no barrier impedes the vortex growth  at phase gradients less than the Landau critical gradient but larger than the inverse core radius,  
there is still a barrier, which blocks phase slips on the very early stage of nucleation of the vortex core. Eventually the barrier for phase slips disappears at the gradient determined by the Landau criterion as usual.

\section{Conclusion: spin superfluidity vs. coherent precession vs. magnon BEC} \label{BEC}

Earlier in the paper we defined  ``spin superfluidity'' directly in terms of an observable effect:
high nearly uniform spin supercurrents transporting spin on macroscopical distances of the order of sample size. However other more formal and abstract definitions of spin superfluidity were suggested. It is difficult to argue about definitions since sometimes it is a matter of semantic taste. Still it is possible to discuss their consistency and rationales.

 \citet{BunV}  identify spin superfluidity with the magnon BEC without paying attention to additional conditions for existence of dissipationless spin transport. They write in the end of Sec. II: ``The magnon BEC is a dynamic state characterized by the Off-Diagonal Long-Range Order (ODLRO), {\em which is the main signature of spin superfluidity}.'' Further, they stress the difference of the non-equilibrium state of coherent precession, which they call magnon BEC, with the equilibrium magnetically ordered system,\cite{Giam} which they do not want to call magnon BEC. They claim that  in the latter system  ODLRO is absent and therefore spin superfluidity must be also absent. According to such an approach  spin supercurrent states in easy-plane antiferromagnets considered above also are not spin-superfluid since they are metastable, i.e., quasi-equilibrium states. 
 
We put aside the question  which magnetic coherent state  can be called magnon BEC and which cannot. Earlier\cite{Adv} we have already  presented our point of view that the term  BEC is not good with respect to magnons in general.  We focus on the suggestion to consider  ODLRO  as a signature of spin superfluidity.  \citet{BunV}  define   ODLRO as  an existence of nonzero average complex quantity $\langle M_x+iM_y\rangle$. This takes place if  the total magnetization precesses. In  the spin-superfluidity example of Sec.~\ref{AF} the total magnetization does not precess, and  $\langle M_x+iM_y\rangle=0$. However, there is a non-zero average complex quantity  $\langle L_x+iL_y\rangle$, where $\bm L$ is the antiferromagnetic vector. One can only guess at why non-zero  $\langle M_x+iM_y\rangle$ is a signature of spin superfluidity but non-zero $\langle L_x+iL_y\rangle$ is not.

 The tendency of identification of coherent spin precession (whether it is called magnon BEC or not) with spin superfluidity becomes a dominant in the latest paper by  \citet{BunLvo}, in which they reported experimental observation of coherent spin precession in easy-plane ferromagnets. This is an interesting result itself, but they presented it also as an evidence of spin superfluidity on the ground that coherent precession must be accompanied by spin supercurrents because of inhomogeneity of samples. As pointed out above, not any  spin supercurrent is a manifestation of spin superfluidity. 
Supercurrents discussed  by  \citet{BunLvo}  transport spin on inhomogeneity scale  in chaotic directions and cannot be an evidence of macroscopical spin transport.  Accepting such a broad definition of spin superfluidity one should consider spin supercurrents in domain walls also as a spin-superfluidity signature. This reduces spin superfluidity to a trivial and hardly interesting effect, making it simply a new fancy name for a well known phenomenon.  

While the  BEC criterion by \citet{BunV} rules out  spin superfluidity in equilibrium magnetically ordered systems despite that stable spin supercurrents are possible there, it predicts spin superfluidity in the coherent state, which was observed by \citet{Dem} in yttrium-iron-garnet films and can be called magnon BEC according to their criterion. However, topology of the order parameter in this case does not allow macroscopic dissipationless spin transport.\cite{Adv}
In summary, the condition for observation of  macroscopic dissipationless spin transport has no connection with the  arbitrary chosen formal  criterion for magnon BEC suggested by \citet{BunV}.

\begin{acknowledgements}
The author appreciates supports by the grant of  the Israel Academy of
Sciences and Humanities and by the EU-FP7 program Microkelvin.
\end{acknowledgements}

\pagebreak

%

\end{document}